\documentclass[reprint,aps,superscriptaddress,prl]{revtex4-1}
\usepackage{graphicx}
\usepackage{dcolumn}
\usepackage{bm}
\usepackage{multirow}
\usepackage[english,russian]{babel}
\usepackage{amsmath}
\usepackage{amsfonts}
\usepackage{amssymb}

\newcommand{\eq}[1]{eq.~(\ref{#1})}

\newcommand{\beq}{\begin{equation}}
\newcommand{\eeq}{\end{equation}}

\newcommand{\la}[1]{\label{#1}}
\newcommand{\bea}{\begin{eqnarray}}
\newcommand{\eea}{\end{eqnarray}}
\newcommand{\ba}{\begin{array}}
\newcommand{\ea}{\end{array}}

\newcommand{\Ro}{{|\cal R}_\omega|^2}
\newcommand{\Ko}{{|\cal K}_\omega|^2}
\newcommand{\Det}{{\mathfrak Det}_{\rm imp}}

\begin{document}

\title{One-dimensional strongly interacting electrons with single impurity: conductance reemergence}
\author{V.V. Afonin}
\affiliation{( A.F.Ioffe Physical-Technical Institute of the RAS,
St.Petersburg, Russia)}
\author{V.Yu. Petrov}
\affiliation{(Petersburg Nuclear Physics Institute, Gatchina 188300, St. Petersburg, Russia)}

\begin{abstract}
We show that conductance of 1D channel with one point-like impurity critically depends on asymptotic behavior of $e-e$ interaction at small momenta $k$
(about inverse length of a channel). Conductance reemerges (contrary to the case of point-like repulsive potential) if potential $V~(k~\!~=~\!~0)~=~0$. For example, this happens
if the bare $e-e$ interaction is screened by the charges in the bulk. The relation of this phenomena to the long-range order present in the Luttinger model is discussed.
 We consider spinless electrons but generalization is straightforward.
\end{abstract}

%

\maketitle

\section{Introduction}

Theory of one-dimensional interacting electrons is under investigation for a long time \cite{T}-\cite{ML}.
Its relativistic analog, two-dimensional QED, also attracted a lot of attention \cite{Sh} in the past, since it is a simplest field theory with confinement. During the time it was understood that one-dimensional pure electronic systems  (in particular, the Luttinger model \cite{L}) are exactly solvable. To date the properties of the clean systems are very well understood. Situation  is different for the 1D channels with some impurities that are understood as short-range barriers with  transition coefficient $K$ and reflection coefficient $R$. The simplest system of this kind (with only one impurity) was considered for the first time in Ref.\cite{FK}.  It turned out that properties of such system depend critically
 on the sign of the electron-electron ($e-e$) interaction. Conductance for  attractive potentials  is equal to the ballistic one and it is not affected by $e-e$ interaction (only the Fermi speed should be renormalized). Conductance for repulsive  potentials  vanishes. These results were obtained in \cite{FK} by the bosonization method.

Another approach with similar results was developed in \cite{M}. The authors returned to the fermion language. Assuming that the interaction is short-range, $V~=~V_0\delta(x)$ and small $V_0~\ll~ 1$, they summed up the leading infrared  logarithms of frequency $(\omega )$ by the  renormalization group method. Next-to-leading corrections to the conductivity were also found in \cite{aristov} by the methods of current algebra.

The approaches of \cite{FK} and \cite{M} have different (but overlapping) regions of applicability. The first approach employs  perturbation theory in reflection (transition) coefficient for an arbitrary attractive (repulsive) potential, while the second one employs  perturbation theory in potential for an arbitrary reflection or transition coefficients. A point-like $e-e$ interaction is assumed in both approaches.

We suggested in \cite{Dual} an alternative approach to the problem based on the path integral formalism. Using the well-known trick \cite{H} we see that  Luttinger model can be interpreted as the system of non-interacting electrons in a random external field. Green functions of  one-dimensional electrons in any external field can be found exactly. We used this fact to construct perturbatively a Green function of the system with impurity. At the end we integrate out
fermions and arrive at a 0+1-dimensional field theory. This theory describes the evolution with time of the electron phase at the point where the impurity is located. It is completely equivalent to the original Luttinger model with one impurity.  For the sake of simplicity  we will consider here only electrons without spin.

Using this theory we were able to prove two theorems. First, conductance of the system is zero (for repulsion) or maximal (for attraction) for a wide class of potentials. The arguments in favor of this statement
for a point-like potential were given earlier in both approaches of \cite{FK} and \cite{M}. We will see below that a necessary condition for such behavior of conductance is that the Fourier transform of the potential $V(k)$ has a non-vanishing limit at $k\to 0$. The second theorem is a general exact property of the theory which one can call {\em duality}. It states that the effective reflection coefficient $\Ro$ in a theory with an attractive potential is equivalent to the effective transition coefficient $\Ko$ in a theory with repulsion if one exchanges $K\leftrightarrow R$ (for a precise formulation of duality transformation of potential, see below). The traces of this property were seen in the perturbation theory in \cite{FK} where duality transformation reduces to  $v_c\!\to \! v_c^{-1}$  ($v_c$ is the renormalized Fermi speed). However, this statement is far more general. It means that it is enough to consider, say, only repulsive potentials.

For a repulsive potential conductance restores if potential vanishes at $k\to 0$. Such a situation takes place in the systems with a small density of carriers when the screening radius is large.  In this case $e-e$ interaction is not point-like from one-dimensional point of view, and it is screened by the image charges on 3-dimensional gates, edges of the channel, etc. Renormalization of the ballistic conductance in such system is finite and will be calculated below. Pay attention that the form and value of conductance is determined not by small $k$ but by the whole region $V(k)$ where the potential is not small. So one needs an approach which is valid for an arbitrary e - e potential, not only for a point-like one as in \cite{FK,M}.

The physical reason for critical phenomena of the conductance in the Luttinger model with an impurity is a long-range order which is present in a system of one-dimensional electrons. It is well-known that its analogue --- the Schwinger model --- exhibits the anomalous breakdown of chiral symmetry. The strength of the interaction in the repulsive Luttinger model is smaller: the system is in the Berezinskii-Kosterlitz-Thouless (BKT) phase\cite{BKT}. Chiral condensate (consisting of pairs of $R$ electron and $L$ hole with finite density) arises only in the limit of an infinitely large interaction. In the case of an attractive potential there is a charged condensate of Cooper pairs with vanishing density (for the channel with infinite length), i.e. one has a BKT phase as well. The Bose-Einstein principle implies that the chiral condensate increases the probability of reflection, i.e. the effective reflection coefficient $\Ro$ at small frequencies, while the charged condensate increases the probability of transition. As a result, $\Ro=1$ for repulsion and $\Ko=1$ for attraction. As we mentioned above, this does not happen if $V(k\to 0)=0$. We will see that in this case the long-range order in the Luttinger model also disappears.  This leads to finite conductance of the channel.

\section{Effective transition/reflection coefficients and conductance}

The Fermi surface in one dimension reduces to two isolated points $\pm p_F$. The electrons with momenta close to the Fermi surface can be divided in right (R) and (L) movers,
\[
\Psi= e^{ip_Fx-i\varepsilon_F t}\psi_R+e^{-ip_F x-i\varepsilon_F t}\psi_L,
\]
where $\psi_{R,L}$ are slowly varying on the scale $1/p_F$.

By means of the Hubbard \cite{H} trick, the Luttinger model  can be reduced to a system of noninteracting electrons in a random external field $U(x,t)$ with a simple Gaussian weight and  subsequent integration in all possible realizations of the field. The Schr\"odinger equation for the non-interaction R,L electrons in the external field reduces to the Dirac equation in $d=1$ (in our units $\hbar=v_F=1$; we will also omit electron charge $e$ to restore it in the final expression for conductivity).
\beq
[i\partial_t\pm i\partial_x -U]\psi_{R,L}=0.
\la{dirac}
\eeq
The Luttinger liquid is a system which can be solved exactly. The ultimate reason for this is  that a one-dimensional fermion Green function in the external field can be found
\bea
G_{R,L}(x,x')&=&G^{(0)}_{R,L}(x-x')e^{i\gamma_{R,L}(x)-i\gamma_{R,L}(x')}
\nonumber\\
\gamma_{R,L}(x)&=&-\int d^2x'G^{(0)}_{R,L}(x,x')U(x'),
\la{green}
\eea
The Green functions $G_{R,L}$ only by phase differs from the free Green functions
\beq
G^{(0)}_{R,L}(x,t)=\frac{1}{2\pi i(t\mp x-i\delta t)}
\eeq
 ($\delta>0$ is infinitesimal.)

A point-like impurity located at $x=0$ mixes left and right electrons. Impurity plays the role of a boundary condition, solutions of \eq{dirac} should be matched at $x=0$. Nevertheless, the general solution in the external field can be found \cite{Dual}. Solution depends on a new functional variable $\alpha(t)$ which is the difference of
phases for $R$- and $L$-electrons at the point of impurity
\beq
\alpha(t)=\gamma_R(0,t)-\gamma_L(0,t).
\eeq
Construction of the Green function with an impurity is impeded  by the Feynman boundary conditions which lead to some integral equation. This equation can be solved  perturbatively either in bare reflection  or in transition   coefficient (for details see \cite{Dual}).

The Luttinger model has  high symmetry: it is invariant
both under gauge (vector) and chiral transformations (the latter symmetry is broken by the anomaly).
The charge density ($\rho=\rho_R+\rho_L$) and current
($j=\rho_R-\rho_L$) can be completely determined from the conservation of  the vector and axial currents:
\beq
\partial_t\rho+\partial_xj=0,
\qquad
\partial_tj+\partial_x\rho=-\frac{1}{\pi}{\partial_x}U+{\mathfrak D}(t)\delta(x).
\la{conserv}
\eeq
Here the first term on the right-hand side is the Adler anomaly \cite{adler}. The second term describes the influence of the impurity, and $\mathfrak D$ is the charge jump at $x=0$ which depends only on phase $\alpha(t)$. It can be calculated if the Green function is known.

Integrating in fermion degrees of freedom allows to present any quantity as a product of Green functions in the external field and fermion determinant describing the sum of the
loop diagrams. As it was shown in \cite{Dual} the effect of impurity is completely determined by the phase $\alpha(t)$: non-trivial part of the Green functions and determinant depends only on $\alpha(t)$. Introducing $\alpha$ as
a new variable one can integrate also in $U(x,t)$ and reduce original $1+1$-dimensional  model with impurity  to the effective $0+1$ field theory (non-local quantum mechanics of the phase $\alpha(t)$).

The conductance of the channel ${\cal C}(\omega)$ is related to the exact transition coefficient
\beq
{\cal C}(\omega) = \frac{e^2\Ko}{2\pi v_r(\omega)}.
\la{conductance}
\eeq
Here $v_r(\omega)$ is the renormalized speed of an electron
\beq
v_r(\omega)=\sqrt{1+\frac{V(\omega)}{\pi}}, \qquad v_c=v_r(0),
\eeq
\beq
\Ro 2\pi\delta(\omega-\omega')=\frac{i\pi}{|\omega| W(\omega)v_r(\omega) }\langle\langle
\alpha(-\omega'){\mathfrak D}(\omega)\rangle\rangle .
\la{obs1}
\eeq
Here the average is understood as an integral with the effective action:
\beq
\langle\langle\ldots\rangle\rangle=
\frac{1}{{\cal Z}}\int\! D\alpha\; \ldots
\Det\exp\left[-\int\!\frac{d\omega}{2\pi}
\frac{\alpha(-\omega)\alpha(\omega)}{2W(\omega)}
\right],
\la{av1}
\eeq
and $W(\omega)$ is
\beq
W(\omega)=-\int\frac{dk}{2\pi i}\frac{4k^2V(k)}{(\omega^2-k^2+i\delta)(\omega^2-v_r^2(k)k^2+i\delta)}.
\la{quadr}
\eeq
The "kinetic energy"  in Eq.(\ref{av1}) is a well-known contribution of the Adler anomaly \cite{Sh} to the effective action rewritten in terms of phase $\alpha(t)$
(see \cite{Dual} for details).

At last, $\Det$ takes care about the loop diagrams describing the multiple rescattering on the impurity. It can be expressed in terms of the charge jump $\mathfrak D$
\bea
\log\Det&=&-\frac{i}{2}\int_0^1\!\! d\lambda\!\int\!\frac{d\omega}{2\pi}\alpha(-\omega){\mathfrak D}[\lambda\alpha](\omega),\nonumber\\
{\mathfrak D}(\omega)&=&2i\frac{\delta}{\delta\alpha(-\omega)}\log\Det[\alpha].
\label{Dlambda}
\eea

Expression (\ref{obs1}) is only one of the possible representations for the effective reflection coefficient. Another useful representation
\beq
\Ro=\frac{2\pi}{|\omega|W^2(\omega)v_r(\omega)}[{\mathfrak g}_0(\omega)-{\mathfrak g}(\omega)],
\la{obs2}
\eeq
relates ${\cal R}_\omega$ to the Green function of the electron phase
${\mathfrak g}(\tau-\tau')=\langle\langle\alpha(\tau)\alpha(\tau')\rangle\rangle$ (${\mathfrak g}_0$ is a Green function without impurity determinant).
Expression (\ref{obs2}) can be obtained from \eq{obs1} taking functional integral by parts, it is one of the Ward identities in the effective theory.

The determinant $\Det$ can be built as a series in bare reflection coefficient
\beq
\log\Det[\alpha]=\sum_{n=1}^\infty \frac{(-1)^{n+1}}{n}\left(\frac{|R|}{|K|}\right)^{2n}{\mathfrak T}_{2n-1}[\alpha],
\la{det1}
\eeq
where
\beq
{\mathfrak T}_{n}\!=\!\int\!\!\frac{d\tau_0\ldots d\tau_n}{(2\pi i)^{n+1}}
\frac{1-\cos[\alpha(\tau_0)-\alpha(\tau_1)+\ldots \alpha(\tau_n)]}
{(\tau_0-\tau_1-i\delta)\ldots(\tau_n-\tau_0-i\delta)} .
\la{bn}
\eeq
In fact, we derived in \cite{Dual} expression not for $\Det$ but for the charge jump ${\mathfrak D}(\omega)$ which is a variational derivative of the determinant in
$\alpha$ according to \eq{Dlambda}.

Formulae  (\ref{obs1})-(\ref{bn}) allow to calculate conductance for attractive e-e interaction as in this case $W(\omega)$ is positive.
For the repulsive interaction one should use a different form of $\Det$. As
it was proved in \cite{Dual}  \eq{det1} can be also presented  in a {\em dual} form as a series in the inverse parameter
\begin{eqnarray}
\log\Det[\alpha]&=&\sum_{n=1}^\infty \frac{(-1)^{n+1}}{n}\left(\frac{|K|}{|R|}\right)^{2n}{\mathfrak T}_{2n-1}[\widetilde{\alpha}]
+\nonumber\\
&&+\int\frac{d\omega}{2\pi}\frac{|\omega|}{4\pi}\widetilde{\alpha}
(-\omega)\widetilde{\alpha}(\omega),
\la{Kexp}
\end{eqnarray}
where $\widetilde{\alpha}(\omega)={\rm sign}(\omega)\alpha(\omega)$. It is natural to call the first term a dual determinant and
combine the second one with the "kinetic energy" \eq{quadr}. Introducing $\widetilde{\alpha}$ as a new variable we obtain a dual
theory with the  kinetic energy
\beq
\widetilde{W}^{-1}(\omega)=-W^{-1}(\omega)-\frac{|\omega|}{2\pi}.
\la{dual-2}
\eeq
Effective coefficients $\Ro$ and $\Ko$ can be written either as functional integrals
\eq{av1} in the original theory or as the functional integrals in the dual theory after the transformation:
\beq
|R|^2\leftrightarrow |K|^2, \quad \Ro\leftrightarrow \Ko, \quad W\leftrightarrow \widetilde{W}.
\eeq

\section{Conductance reemergence}

It is well-known \cite{FK} that the coefficient $\Ro\to 0$ for an attractive potential and $\Ko\to 0$ for a repulsive one when  $\omega\to 0$.
This happens  only if $V(k\!\!=\!\!0)\!~\neq\!~0$ due to the infrared divergency. We reproduced this result for our effective 0+1 dimensional theory in \cite{Dual}.

An example of a different behavior
(for a repulsive potential with $V(k\!\!=\!\!0)=0$) is given by a system of one-dimensional electrons with small concentration. We will see that conductance remains finite in this case.

Indeed, if the concentration is small then  the 3D screening radius can be much larger than the width $d$ of the  channel (which is considered to be zero in our one-dimensional theory). The bare $e-e$ interaction is screened by the "image charges"\; that arise on the split gates. The screened interaction is Coulomb one at distances smaller than the distance to the gate ($l$) and it is dipole-dipole in the opposite limit \cite{Gur}:
\beq
V(k) = -\zeta\left\{
\begin{array}{lc}
\log|k|d, & kl\gg 1 \\
(kl)^2\log|k|d, & kl\ll 1
\end{array}
\right..
\la{pot}
\eeq
Here $\zeta\equiv 2/\pi a_Bp_F\gg 1$, $a_B$ is the Bohr radius. We assume that $d\ll l\ll L$ where $L$ is the length of the one-dimensional channel and $\zeta$ is a largest parameter of the problem. Moreover, we will consider the case when transition coefficient is small: $|K|^2\ll 1$. Then one can leave only first term in the expansion (\ref{Kexp}) and calculate the charge jump $\mathfrak D$ according to \eq{Dlambda}. Taking Gaussian integral in $\alpha$ we arrive at
\beq
\Ko=\frac{|K|^2}{\pi }\int d\tau\frac{1-\cos\omega\tau}{|\omega|\tau^2}e^{-\sigma(\tau)},
\la{ATTR}
\eeq
where
\beq
\sigma(\tau)=\int\frac{d\omega'}{2\pi}\widetilde{W}
(\omega')\left(1-\cos\omega'\tau\right).
\la{XI}
\eeq
Behavior of transition coefficient at small $\omega$ is related to the asymptotic $\sigma(\tau)$ at large $\tau$. The  kinetic energy  $\widetilde{W}(\omega)$ in \eq{dual-2} is nonsingular  for repulsive potential (\ref{pot}) at $\omega\to 0$. For this reason $\sigma(\tau)$ has a finite limit $\sigma_\infty$ at $\tau\to\infty$. This limit determines
the conductance
\beq
{\cal C}(\omega=0) = \frac{e^2|K^2|}{2\pi}e^{-\sigma_\infty}, \qquad \sigma_\infty=\int\!\frac{d\omega}{2\pi}\widetilde{W}(\omega)
\la{finite-1}
\eeq
The renormalized Fermi speed $v_c=1$ here.

At $l\sqrt{ \zeta }/ d\gg 1$ the  main contribution to the integral in \eq{finite-1} comes from $kl~\gg~1$, where the potential is not screened. The kinetic energy
$\widetilde{W}(\omega)$ is determined by the pole $|\omega|=v_r(k_0)|k_0|\approx |k_0|\sqrt{\zeta\log{(1/k_0d)}}$ in \eq{quadr}. It is equal to
\begin{equation}
\widetilde{W}(\omega ')=\frac{2\pi}{|\omega '|}\sqrt{\zeta
\log{(\sqrt\zeta /|\omega '|d )}}
\label{W}
\end{equation}
at $\omega ' l\ge 1 $. Finally
\beq
\sigma_\infty\approx \frac{4}{3}\sqrt\zeta
\left(\log\sqrt\zeta\frac{l}{ d}\right)^{\frac{3}{2}}\gg 1
 \eeq
Hence, conductance of channel in this limit is small. Let us note that conductance is determined not by the $k\to 0$ but by $k\sim 1/l$
where interaction of electrons is important.

Consider now also the opposite limit $\sigma_\infty\ll 1$ while $K$ is not necessarily small. It can be implemented at intermediate concentrations if a 3-dimensional screening radius is of the order of the channel thickness. In this case we are dealing with a weak interaction and the conductance is determined by an expansion in powers of $\alpha$
\bea
{\mathfrak T}_{2n-1}&=&\frac{n}{2!}\int\!\frac{d\omega}{(2\pi)^2}|\omega|\alpha(\omega)\alpha(-\omega)-
\frac{n^2}{4!}\int\!\frac{d\omega_1\ldots d\omega_4}{(2\pi)^4}\times\nonumber\\
&&\times\delta\left(\sum_{i=1}^4\omega_i\right)
\Gamma_4(\omega_i)\alpha(\omega_1)\ldots\alpha(\omega_4)+\ldots,
\eea
where the vertex $\Gamma_4$ is
\beq
\Gamma_4(\omega_i)=\sum_i|\omega_i|-\frac{1}{2}\sum_{i< j}|\omega_i+\omega_j|
.\eeq
This vertex is equal zero if any frequency $\omega_i$ vanishes. Substituting this expression into \eq{det1} and taking an integral over $\alpha$ with
the quadratic form $\widetilde W$ we obtain the effective transition coefficient
\beq
\Ko=|K|^2-|R|^2|K|^2\sigma_\infty+O(\sigma_\infty^2),
\la{smalls}
\eeq
which again means  that conductance is non-zero according to \eq{conductance}. Notice that this expression obeys duality which requires that
the term linear in interaction should be symmetrical under $K\leftrightarrow R$ exchange. At small $|K|^2$ we return here to the expression in \eq{finite-1} expanded at small $\sigma_\infty$ \footnote{For a point-like potential one has to substitute here $\sigma_\infty=2{\tilde \nu}\log(M/\omega)$. Then \eq{smalls} will coincide with the expression for the first logarithmic correction to conductance obtained in \cite{M}. This expression was the base for the renorm-group approach developed in Ref. \cite{M}. Comparison of the results obtained in our effective field theory with the renorm-group considerations will be published elsewhere.}.

The physical reason for restoration of the conductance is, in fact, disappearance of long-range order in the  system. In fact, one of two continuum symmetries present in the Luttinger model (chiral and gauge invariance) is always broken. To prove this let us consider first the case of repulsive potential. To check, whether the chiral order is present in the system we consider the following correlator
\beq
{\cal G}(R)=\langle\psi_R(R,0)\psi^+_L(R,0)\psi^+_R(0,0)\psi_L(0,0)
\rangle.
\la{chiral-cor}
\eeq
There are a number of methods to calculate (\ref{chiral-cor}) but the simplest is to use again the Hubbard trick \cite{H}. Then one can present ${\cal G}(R)$ as a product of two
Green functions $G_{R,L}$ of electron in the external field (\ref{green})%
\begin{equation}
{\mathfrak G}(x,\bar{x},y,\bar{y})= \overline{G_R(x,\bar{x})G_L(y,\bar{y})}.
\end{equation}
where averaging is a functional integral
\begin{eqnarray}
&&\frac{1}{{\cal Z}}\int\! DU\ldots \exp\left\{\frac{i}{2}\int\frac{d^2k}{(2\pi)^2}
\frac{U(-k,-\omega)U(k,\omega)}{V(k)}\times
\right. \nonumber\\
&&
\left.\times
\frac{\omega^2-v_r^2(k)k^2+i
\delta}
{\omega^2-k^2+i\delta}\right\},
\label{U}
\end{eqnarray}
 and $x=(x,t)$ is two-dimensional coordinate (for details see Appendix of Ref.\cite{F1}).

According to \eq{green} $R$- and $L$- electrons acquire a phase in the external field which is linear in $U$. Therefore integral in $U$ is gaussian
$$
{\mathfrak G}_{RL}(x,\bar{x},y,\bar{y})=G^{(0)}_{R}(x,\bar{x})G^{(0)}_{L}
(y,\bar{y})\cdot$$
\beq
\cdot\exp\left\{
-\frac{i}{2}\int\frac{d^2k}{(2\pi)^2}\frac{V(k)}
{(\omega^2-v_r^2(k)k^2+i\delta)(\omega^2-k^2+i\delta)}\right\}.
\label{ex1}
\eeq
$$.\left\{\left|
(\omega+k)\left(e^{-ik\cdot x}-e^{-ik\cdot \bar{x}}\right)+
(\omega-k)\left(e^{-ik\cdot y}-e^{-ik\cdot \bar{y}}\right)
\right|^2
\right\}$$

Now, to investigate chiral properties of the system we  put here ${x}=\bar{y}=R$ and $\bar{x}=y=0$ and take  the limit $R\to\infty$,
\begin{eqnarray}
{\cal G}(R)&=& \frac{1}{4\pi^2R^2}\exp\left[\int\!\frac{d^2k}{(2\pi)^2i}
|1-e^{ik\cdot R}|^2\times
\right.\nonumber\\
&&\times\left.\frac{2k^2V(k)}{(\omega^2-(v_r)^2k^2+i\delta)
(\omega^2-k^2+i\delta)}
\right].
\la{GR}
\end{eqnarray}
The asymptotic at large $R$ of this expression is given by:
\beq
{\cal G}(R)=\frac{1}{4\pi^2R^2}
\exp[-2\int^{\infty}_{1/R}\frac{dk}{|k|}\left(
v_r(k)^{-1}-1\right)] ,
\la{chiral}
\eeq
(Convergence of the integral in the ultraviolet is provided by the condition $v_r(k)\to 1$ at $k\to\infty$).
The renormalized Fermi speed $v_r(k)>1$ for all $k$. For the point-like interaction Eq.(\ref{chiral}) turns into
\beq
{\cal G}(R)\sim 1/R^{2-2\eta}, \quad \eta=1-\frac{1}{v_c}
\eeq
where $v_c$ is the speed on the Fermi surface. In the limit $v_c\to\infty$ (an infinitely strong interaction)  ${\cal G}(R)$  goes to some constant at large $R$. It means that  a chiral condensate $\langle\psi_R\psi^+_L\rangle\neq 0$ is formed  in the system. This phenomenon is known to happen also in the Schwinger model (see, e.g., \cite{smilga}).
For a finite interaction there is no condensate but the correlator ${\cal G(R)}$ decays slower than for free electrons and long-range order still exists.
The number of correlated $R\bar{L}$ pairs is macroscopically large $\sim L^{\eta}$ and the system is in the BKT phase.
We have constructed an exact wave function of the ground state of the Luttinger model and investigated the nature of this
phase in Ref.\cite{KT}.

The macroscopic number of $R\bar{L}$ pairs in the vacuum of Luttinger model amplifies  the back-scattering of electrons on impurity owing to Bose-Einstein principle. As a result, as it is well-known \cite{FK}, transition coefficient tends to zero at $\omega\to 0$:
\beq
|K|^2 \sim \omega^{2\eta}
\eeq

For the dipole-dipole interaction (\ref{pot}) the integral (\ref{chiral}) is infrared convergent since $v_r(k)=1$ at $k\leq 1/l$.  In this case at large distances
$$
{\cal G}(R)\sim 1/R^2,
$$
is the same as for free particles. Only  pairs  with momenta $k \ge 1/l$ are correlated strongly and the number of such pairs does not increase  with  system volume $L$.  As a result transition coefficient remains finite.

Similar situation takes place for an attractive interaction but for the {\em charged} condensate (as for superconductivity). To reveal this condensate one considers the correlator:
\beq
\widetilde{{\cal G}}(R) =
\langle\psi^+_R(R,0)\psi^+_L(R,0)\psi_R(0,0)\psi_L(0,0)\rangle  ,
\eeq i.e. we have to put ${x}=y=R$ and $\bar{x}=\bar{y}=0$.
In the same way one has
\begin{eqnarray}
\widetilde{{\cal G}}(R)&=& \frac{1}{4\pi^2R^2}\exp\left[\int\!\frac{d^2k}{(2\pi)^2i}
|1-e^{ik\cdot R}|^2\times
\right.\\
&&\left.\frac{2\omega^2V(k)}{(\omega^2-(v_r)^2\omega^2+i\delta)
(\omega^2-k^2+i\delta)}
\right]
\la{GA}
\end{eqnarray}
or
\beq
\widetilde{{\cal G}}(R)=\frac{1}{4\pi^2R^2}
\exp[-2\int_{1/R}\frac{dk}{|k|}\left(
v_r(k)-1\right)] .
\la{chiral2}
\eeq
For a point-like potential superconducting condensate arises for $v_c=0$, while for $0<v_c<1$ the system is in the BKT phase:
\beq
\widetilde{{\cal G}}(R)\sim 1/R^{2-2\widetilde\eta}, \qquad \widetilde{\eta}=v_c-1
\eeq
and reflection coefficient $|R|^2\sim \omega^{2\widetilde{\eta}}$. For the screened potential BKT phase disappears and renormalization of conductance by interaction is finite.

We see that in all cases properties of  ${\cal G}(R)$ at large $R$ and of the conductance for small $\omega$  coincide for arbitrary e-e potential. Thus,  disappearance or restoration of the conductance is related to the long-range order in the Luttinger model.

\begin{acknowledgements}
The authors are grateful to M.Eides for very useful discussions.
This paper is supported in part by RFBR grant {\bf 15-02-01575-A.}
The work of V.P. is supported by Russian Scientific Foundation, grant
{\bf 14-22-0281}
\end{acknowledgements}

\end{document}